\documentstyle[prd,aps,
preprint,
tighten,
multicol,axodraw]{revtex}

\begin{document}

\draft
\preprint{IFT-99/30 \hspace{.5ex} KIAS-P99107 \hspace{.5ex} hep-ph/9912210}

\title{Slepton Flavour Mixing and Neutrino Masses}
\author{ Eung Jin Chun$^a$ and Stefan Pokorski$^b$ }
\address{ $^a$Korea Institute for Advanced Study, 
           Seoul 130-012, Korea \\
           $^b$Institute of Theoretical Physics,
           Warsaw University, Poland}

\maketitle

\begin{abstract}
It is pointed out that important contribution to neutrino masses and
mixing can come from flavour violating slepton exchange.
For low and intermediate $\tan\beta$ such effects can dominate
over the usual tau Yukawa coupling effects included 
in the renormalization group evolution.
The mixing angles satisfy then the relation different 
from the fixed point solution of the renormalization group equation, 
and the desired neutrino mass splitting can be generated.
\end{abstract}


%\begin{multicols}{2}

Observation of atmospheric \cite{skatm} and solar neutrinos
\cite{solars} provides important indications that neutrinos
oscillate between different mass eigenstates. If we leave out the
not yet confirmed LSND result \cite{lsnd}, the atmospheric and
solar neutrino oscillations can be explained in terms of three
known flavours of neutrinos. We have then $|\Delta m^2|\equiv
|m_2^2-m_1^2| \approx 5\times10^{-5}$, $5\times10^{-6}$ or
$10^{-10}\, eV^2$ for the large (LAMSW), small angle matter
conversion (SAMSW), or vacuum oscillation (VO) solution to the
solar neutrino problem, respectively, and  $|\Delta M^2| \equiv
|m_3^2-m_2^2| \approx 5\times10^{-3}\, eV^2$ for the atmospheric
neutrino oscillation.  Moreover, combining the information on the
solar and atmospheric neutrino mixing angles
%$\sin^22\theta_{sol} \sim 10^{-2}$ (SAMSW), $1$ (LAMSW or VO),
%and $\sin^22\theta_{atm} \sim 1$,
with non-observation of the disappearance of $\bar{\nu}_e$ in the
reactor experiment \cite{chooz}, for each solution (LAMSW, SAMSW,
VO) to the solar neutrino problem we can infer the gross pattern
of the 3x3 neutrino mixing matrix
\begin{equation}
 U=\pmatrix{ c_2c_3 & s_3c_2 & s_2 \cr
            -c_1s_3-s_1s_2c_3 & c_1c_3-s_1s_2s_3 & s_1c_2 \cr
         s_1s_3-c_1s_2c_3 & -s_1c_3-c_1s_2s_3 & c_1c_2 \cr }
\end{equation}
with $\sin^22\theta_1\approx\sin^22\theta_{atm} >0.82$ and
$U_{13}^2=s_2^2 < 0.2$. Another important experimental information
comes from non-observation of neutrinoless double $\beta$
decay \cite{0nbb}, which provides a bound on the ($ee$) component of
the neutrino mass matrix, that is,
\begin{equation} \label{Mee}
 {\cal M}_{ee} =|m_1c_2^2c_3^2+m_2c_2^2s_3^2+m_3s_2^2| < 0.2  eV \,.
\end{equation}

The question of theoretical explanation of neutrino masses and
mixing has been discussed in a large number of papers \cite{revs}.
The measured neutrino mass matrix is linked to the fundamental
mass generation by two steps.  The first one is some flavour
violation in the lepton sector which may have origin at a certain
fundamental scale.  The second step consists of including quantum
corrections to obtain physical neutrino masses. The leading
effects can be taken into account by the renormalization group
(RG) evolution from a fundamental scale $M$ to the weak scale
$M_Z$.  An interesting question is the stability of various mass
textures assumed at $M$ with respect to quantum corrections.
Recently, there have been many papers dealing with this issue
\cite{rge1,rge2,casas,barbi,chan} in the context of the
see-saw mechanism which generates tiny neutrino masses 
%through an effective dimension five operator,
% ${\cal L} = - C^{ab} (Hl_a) (Hl_b)/M + h.c. $.
below the Majorana scale $M$.

In this letter, we point out that, in addition to the corrections
due to lepton Yukawa couplings usually included in the RG
evolution and already encoded in the tree-level neutrino mass
matrix at the weak scale, important
contribution can come from flavour changing slepton exchange. We
estimate such effects in one-loop approximation and then discuss
their potential impact on the neutrino mass and mixing patterns
at the scale $M_Z$.

\smallskip

The one-loop corrected neutrino mass matrix is 
\begin{equation} \label{mpole}
 M^{\nu}_{ab} = m_a \delta_{ab} + {1\over2}(m_a+m_b)I_{ab} \,.
\end{equation}
Here $I_{ab}$'s (defined in the tree-level mass eigenstate basis)
are related to the quantities $I^w_{ij}$ (defined in the
flavour basis) as follows;
\begin{equation} \label{Iai}
 I_{ab} = \sum_{ij} I^w_{ij} U_{ia} U_{jb} \,.
\end{equation}
Note that, in the flavour basis, charged lepton masses are
diagonalized but sleptons may have flavour changing masses.  We
calculate $I^w_{ij}$ coming from slepton flavour mixing.  There
are diagrams involving only  left-handed (LL)  or
right-handed (RR) slepton mixing, and left-right mixing (LR). The
latter two involve charged lepton Yukawa couplings and thus become
sizable for large $\tan\beta$. For the order of magnitude
discussion in this letter, we estimate only (LL) type [see Fig.~1]
adopting the mass insertion method as in Ref.~\cite{gabbi}. 
For this, one introduces
$\delta^l_{ij}\equiv\Delta_{ij}/\tilde{m}_i\tilde{m}_j$ for $i\neq
j$ where $\tilde{m}^2_i$ and $\Delta_{ij}$ are the diagonal and
off-diagonal components of slepton mass matrix in the flavour
basis, respectively.  
We then have
\begin{eqnarray}
 I_{ii}^w &=& {g^2\over16\pi^2}
          B_1(p^2,\tilde{m}_i^2,m^2_{\tilde{W}^-})\,,  \nonumber\\
 I_{ij}^w &=& {g^2\over16\pi^2} \delta^l_{ij}
          C(\tilde{m}_i^2,\tilde{m}_j^2,m^2_{\tilde{W}^-}) \,,
\end{eqnarray}
where $B_1(p^2,m_1^2,m_2^2)=1/2\hat{\epsilon}- \int_0^1dx\, x \ln[xm_1^2+
(1-x) m_2^2-x(1-x)p^2]/Q^2$, and
$C(m_1^2,m_2^2,m_3^2)=m_1m_2[B_1(p^2,m_1^2,m_3^2)-B_1(p^2,m_2^2,m_3^2)]/
[m_1^2-m_2^2]$.
Note that we treated the chargino $\tilde{W}^-$ as a mass
eigenstate for the sake of simplicity and there is also a similar
contribution from the zino exchange which we neglect.

\smallskip

The first question we can address is what are the upper bounds on
flavour off-diagonal slepton mass matrix elements ({\it
i.e.}~$\delta^l_{ij}$) which follow from the requirement that our
new contribution is consistent with the observed neutrino mass matrix 
at the scale $M_Z$, without the necessity of large cancellations
among different contributions.
To this end, let us recall that the
following low energy mass patterns are consistent with the
experimental data:

(i) $ m_1^2 \approx m_2^2 \ll m_3^2 \approx \Delta M^2$ (hierarchical )

(ii) $m_1^2 \approx m_2^2 \approx \Delta M^2 \gg m_3^2$ (inverse
hierarchical)

(iii) $m_1^2 \approx m_2^2 \approx m_3^2 \gtrsim \Delta M^2$ (full degeneracy)

\noindent Here we have taken the patterns with partial or full
degeneracy since the fully hierarchical pattern gets essentially
no change as can be inferred from Eq.~(\ref{mpole}). In the
leading approximation (up to terms ${\cal O}(I^2_{ab})$, $a \neq b$), 
the one-loop correction to the mass eigenvalues
is given by $m_a \to m_a(1+I_{aa})$ and thus 
%the requirement of no large cancellations between different
%contributions gives the bounds
we get the bounds
\begin{equation} \label{Ibound}
  |I_{11} - I_{22}| \lesssim  {\Delta m^2 \over 2m_1^2}\,,\quad
  |I_{22} - I_{33}| \lesssim  {\Delta M^2 \over 2m_3^2}\,,
\end{equation}
where the second inequality  applies only for pattern (iii).
Taking one insertion $I^w_{ij}$ at a time,
degenerate uncorrected neutrino masses and degenerate diagonal
slepton mass matrix elements, and without resorting to
cancellations among the mixing elements $U_{ab}$, 
Eqs.~(\ref{Iai},\ref{Ibound}) give the bound on the 
chosen $I^w_{ij}$.  For LAMSW or SAMSW values of 
$\Delta m^2$, in case (i) with 
$10^{-5}eV^2\lesssim m^2_1\lesssim 10^{-4}eV^2$ 
we get $I^w_{ij}\lesssim 1$. Since
$I^w_{ij}\lesssim (10^{-2}-10^{-3})\delta^l_{ij}$, we get no bounds
on $\delta^l_{ij}$'s. Similar conclusion holds  in case (ii).
In case  (iii) we get the bound
\begin{equation}\label{MSWbound}
\delta^l_{ij}\lesssim (10^{-2}-10^{-3})\frac{1}{m_1^2/eV^2} \,.
\end{equation}
%which gives $\delta^l_{ij}\lesssim (10^{-2}-10^{-3})$ for $m_1=1eV$.
In order for  the VO solution to be realized,
severe bounds have to be fulfilled;
\begin{equation} \label{VObound}
 \delta^l_{ij} \lesssim (10^{-7}-10^{-8})
\frac{1}{m_1^2/eV^2}
\end{equation}
for (i, ii) or (iii).  These bounds have to be contrasted with the
present experimental limit on the flavour changing masses coming
from $\mu \to e\gamma$ decay which reads $\delta^l_{12} \lesssim
10^{-2} (\tilde{m}/100  GeV)^2$ \cite{gabbi,barbi95}.
Thus, our considerations put for some of the textures much
stronger bounds, which are insensitive to the values of
$\tilde{m}_i$.

There are also limits on the splittings among the diagonal
masses.  For instance, if $\tilde{m}_1 = \tilde{m}_2 \neq
\tilde{m_3}$, we have $I^w_{11} = I^w_{22} \neq I^w_{33}$. Taking
all $\delta^l_{ij}=0$, we get $I_{11}-I_{22}=(I^w_{33}-I^w_{11})
(U_{31}^2-U_{32}^2)$. This puts the same bounds as in
Eqs.~(\ref{MSWbound},\ref{VObound}) to the quantity,
\begin{equation} \label{diagbound}
%&& (U_{31}^2-U_{32}^2)
% [B_1(p^2,\tilde{m}_3^2,m_{\tilde{W^-}}^2) -
%   B_1(p^2,\tilde{m}_1^2,m_{\tilde{W^-}}^2)]  \nonumber\\
%&& \approx 
(U_{31}^2-U_{32}^2) \ln(\tilde{m}_1 /\tilde{m}_3)
\end{equation}
assuming $\tilde{m}_i \gg m_{\tilde{W}^-}$. Note that this quantity
vanishes at the point $U_{31}^2=U_{32}^2$,
like the tau Yukawa coupling contribution  \cite{barbi,chan}. 
Otherwise, high degeneracy among the diagonal slepton masses is
required when the VO solution with a degenerate mass pattern
is realized in  supersymmetric models.

\smallskip

The next question we ask is  whether flavour mixing in the
slepton sector can play a positive role in generating the desired
neutrino mass splitting and mixing angles with degenerate
textures assumed at the scale $M$. From the discussion above one
sees that indeed there is some room for that. Such effects come on
top of the usual tau Yukawa coupling effects included in the
RG evolution, which contribute to the mass-squared difference
$\Delta m_{ij}^2\approx{\cal O}(m^2_i\epsilon_\tau)$ where
$\epsilon_\tau=h^2_\tau \ln(M/M_Z)/16\pi^2\approx
10^{-5}\tan^2\beta$. For $\tan\beta$ in the range ($1-50$), 
$\epsilon_\tau$ is in the range $(10^{-5}-10^{-2})$ and the
equation $\Delta m^2_{ij}\approx{\cal O}(m^2_i I^w_{kl})$ tells us
that, generically, the slepton exchange effects can be comparable
or even dominant for $I^w_{ij} \gtrsim 10^{-5}$. For the
off-diagonal elements, this requires $\delta^l_{ij}\gtrsim {\cal
O}(10^{-2})$ which is possible for $\delta^l_{13}$ and $\delta^l_{23}$
but excluded for $\delta^l_{12}$ by the bound from $\mu\rightarrow
e\gamma$ decay. On the other side, since realistically
$\delta^l_{ij}<1$, the flavour changing slepton exchange effects can
be important only for $\tan^2\beta<{\cal O}(100)$. The diagonal
$I^w_{ii}$ can be important even for a $10\%$ violation of
universality of the diagonal slepton masses. In addition, those
generic orders of magnitude can be modified in special cases when
e.g.\ the leading Yukawa coupling effects cancel out due to
special relations between the mixing angles.

Let us now discuss the three mass patterns in turn. For obtaining
LAMSW or SAMSW solution with pattern (i) or (ii), the slepton
loop corrections could be important only for maximal flavour
mixing $\delta^l_{ij}\approx{\cal O}(1)$. Although not yet excluded
for $\delta^l_{13}$ and $\delta^l_{23}$, it looks unrealistic. We
expect therefore that in those cases the conclusions based on the
usual tau Yukawa coupling effects included in the RG evolution
remain unaltered and we focus our discussion on the VO solution
and on pattern (iii) in general.
For the VO solution with (i) or (ii) we consider two cases.

$\bullet$ $m_1=m_2$: We recall that this initial condition
places us at the scale M in the infrared  fixed point $U_{31}=0$ or
$U_{32}=0$ \cite{casas,chan}. At the fixed point, small solar mixing
$\theta_3\ll1$ requires the smallness of $s_2\ll1$, and large
mixing is  consistent with a rather large value of $s_2^2\lesssim
0.2$ \cite{chan}. Since $\Delta m^2_{12}\approx
m^2_1\epsilon_\tau\approx (10^{-5}\tan^2\beta)m_1^2$, the VO
solution can be obtained  only for $10^{-8}eV^2\lesssim
m^2_1\lesssim 10^{-5}eV^2$ (which falls into pattern (i)) and
for sizable range of $\tan\beta$, with larger values corresponding
to smaller $m^2_1$. Thus, for low $\tan\beta$ and small enough
$m^2_1$ there is room for important contribution $\Delta
m^2_{12}=4I_{12}m^2_1$. For instance, for $m_1^2\approx 10^{-7}
eV^2$ and $\tan^2\beta\approx{\cal O}(10)$, $\epsilon_\tau$ is
one order of magnitude too small to obtain the VO solution but the
latter can be obtained with $\delta^l_{ij}\approx {\cal O}(0.1)$.

For such a scenario, it is very interesting to discuss the impact
of the slepton loop corrections on the mixing angles which, as we
said above, must satisfy one of the relations $U_{31}=0$ or
$U_{32}=0$ at the scale $M$. For that, we consider the
rediagonalization of the one-loop corrected mass matrix
(\ref{mpole}) assuming that, for instance, $I^w_{23} \gg \epsilon_\tau$ 
and the other $I^w_{ij}$ are negligible. Also,
we recall that in the considered case, at the tree-level at $M_Z$,
$m^2_1\approx m^2_2 \ll m^2_3\approx 10^{-3}eV^2$ and
$U_{31}=0$. We get $I_{12}=2I^w_{23}U_{21}U_{32}$,
$M^{\nu}_{22}-M^{\nu}_{11}\approx
m_1(I_{22}-I_{11})\approx 2m_1 I^w_{23}U_{22}U_{32}$. Moreover, since
$m_{1,2}\ll m_3$, we can make a seesaw diagonalization separating
the first 2x2 block and (33) component and treat $M^\nu_{13},
M^\nu_{23}$ as  small corrections mixing these two blocks.  Then,
the effective contribution to the first 2x2 block is of the order
$m_3 I^2_{13,23}$ which is smaller than the leading term $m_1
I_{12}$ as we generically have  $m_1 \gtrsim m_3 I_{ab}$ for our case
of interest.
%$m_1 \gtrsim m_3 I \sim m_3 \times 10^{-3} eV$. This is
%generically true for our case of interest.
The rediagonalization of the 2x2 block of $M^\nu$ is performed by
a rotation $V(\theta^\prime_3)$. The final neutrino mixing matrix
reads
\begin{equation}\label{finalU}
{\tilde U}({\tilde\theta}_i)=U(\theta_i)V(\theta^\prime_i)
\end{equation}
where ${\tilde\theta}_1=\theta_1$, ${\tilde\theta}_2=\theta_2$,
${\tilde\theta}_3=\theta_3+\theta^\prime_3$. We can immediately
conclude (or check by explicit calculation remembering that the
angles $\theta_i$ satisfy the relation $U_{31}=0$) that the final
angles ${\tilde\theta}_i$ satisfy the relation corresponding to
\begin{equation} \label{fx1}
  I_{12}=I^w_{ij}({\tilde U}_{i1}{\tilde U}_{j2}+{\tilde U}_{j1}{\tilde U}
  _{i2})=0\,
\end{equation}
with $I^w_{23} \neq 0$ and vanishing other $I^w_{ij}$'s. This
relation reads
\begin{equation}\label{finalangle}
\tan 2{\tilde\theta}_3\tan 2{\tilde\theta}_1=\frac{2{\tilde
s}_2}{1+{\tilde s}^2_2} \,.
\end{equation}
Thus, the angles ${\tilde\theta}_1$ and ${\tilde\theta}_2$ are
fixed by the initial conditions \cite{chan} 
and the angle ${\tilde\theta}_3$
is modified by the slepton loop corrections to satisfy
Eq.~(\ref{finalangle}). We observe that, similarly to the RG fixed
point relation $U_{31}=0$, the final relation (\ref{finalangle})
implies very small solar mixing for ${\tilde s}^2_2 \ll 1$.

It is straightforward to extend our consideration to the case of
dominant contributions from $I^w_{13}$ (the possibility of a
dominant $I^w_{12}$ is excluded as discussed).
%by the bound from $\mu\to e\gamma$ decay). 
The resulting
relations for the angles can be read off from Eq.~(\ref{fx1}) and
are collected in the left column of Table I. One can see that the
limit $s_2^2 \to 0$ is consistent with the maximal solar mixing
angle $\sin^22\theta_3 \to 1$ for dominant $I^w_{13}$.
The same results follow for pattern (ii).

$\bullet$ $m_1=-m_2$:   The mixing angles are stable  \cite{casas,chan}
under all loop corrections as
can be inferred from  $M^{\nu}_{12}={1\over2}(m_1+m_2)I_{12}=0$.
That is, all mixing angles are determined initially at $M$. Apart
from the different behavior of mixing angles, the previous
conclusions for $\Delta m^2$  hold here as well. One additional
aspect is  that for pattern (ii) the  $\epsilon_\tau$ contribution  
can be suppressed  when $U_{31}^2=U_{32}^2$, 
a special case of which is the bimaximal mixing solution 
with $s_2=0$. 
The next-to-leading contribution of $\epsilon_\tau^2$ is also absent
under this  condition \cite{chan}.
%$U_{31}^2=U_{32}^2$\cite{chan}. 
It is interesting to observe that the desired value for $\Delta m^2$ of
the VO solution can, however,  come from a very tiny slepton
flavour mixing in  $I^w_{ij}$.

\smallskip

The most interesting case is the fully degenerate pattern (iii).
An attractive feature of this pattern is the possibility of
the cosmologically interesting mass range, $m_3\sim 1 eV$. With
only the tau Yukawa coupling effects, one cannot explain two
distinct mass-squared differences. The slepton exchange
contributions read
\begin{eqnarray}
\Delta m^2 &\approx& 4m_1^2 I^w_{ij}(U_{i1}U_{j1}-U_{i2}U_{j2})\,,
 \nonumber\\
\Delta M^2 &\approx& 4m_1^2 I^w_{ij}(U_{i2}U_{j2}-U_{i3}U_{j3})\,,
\end{eqnarray}
and it is interesting to check whether they can help to obtain the
correct mass-squared differences.
There are two realistic possibilities for $|m_1|=|m_2|=m_3$.

$\bullet$ $m_1=m_2=-m_3$: The slepton exchange modifies the
mixing angles and gives the relation corresponding to
Eq.~(\ref{fx1}), however, it does not modify the
RG evolution result $\Delta m^2 \sim \Delta M^2$.

$\bullet$  $m_1=-m_2=m_3$: This is the case where the mixing
angle relation corresponding to $I_{13}=0$ [see  Table I] is
realized. As above, the $\epsilon_\tau$ or $I^w_{ij}$ dominance
leads generically to $\Delta m^2 \sim \Delta M^2$. However, there
is a texture to accommodate both $\Delta m^2$ and $\Delta M^2$.
This is the one with bimaximal mixing
%$\sin^22\theta_1=\sin^22\theta_3=1$ (consistent with
%Eq.~(\ref{Mee})), 
and the $I^w_{23}$ dominance.
As shown in Table I, this leads to $s_2=0$ at which
$U_{31}^2=U_{32}^2$ as well as $U_{21}U_{31}=U_{22}U_{32}$, and
thus $\Delta m^2$ has neither $\epsilon_\tau$ nor $I^w_{23}$
contribution. As a consequence, we have
\begin{equation}
 \Delta M^2 \sim m_1^2 I^w_{23}\,,\quad
 \Delta m^2 \sim m_1^2 (I^w_{12}\,, I^w_{13})\,,
\end{equation}
implying that the bimaximal mixing  solution can be obtained with
$I^w_{23} \sim 10^{-3}eV^2/m^2_1$, and $I^w_{12},I^w_{13} \sim
10^{-5}eV^2/m^2_1$ ($10^{-10}eV^2/m^2_1$) for the LAMSW (VO) solution.
Similar conclusions can be drawn with $-m_1=m_2=m_3$. 
The corresponding mixing angle relations (obtained from $I_{23}=0$) 
are listed in Table I.

Let us finally mention the case with dominant diagonal $I^w_{ii}$.
If $I^w_{33}\neq I^w_{11}=I^w_{22}$ as in Eq.~(\ref{diagbound}), 
there follows the same behavior as with dominant $\epsilon_\tau$. 
For instance, one has $I_{12}= (I^w_{33}-I^w_{11}) U_{31}U_{32}$ and thus
the mixing angle relation $U_{31}=0$ or $U_{32}=0$ 
fixed by the RG evolution (with $m_1=m_2$) is not altered.  

\smallskip

So far, we discussed the role of loop corrections due to flavour
violation in the slepton sector in quite general terms, based on
the fact that current experiments provide no serious constraints
on $I^w_{ij}$ except $I^w_{12}$.  Indeed, our discussion to be
more predictive, requires some theoretical framework for the
soft sfermion masses departing from universality.
As an example we take a supersymmetric SO(10) model
with universal soft terms assumed at the Planck scale \cite{barbi95}.
In this framework, the large top (neutrino) 
Yukawa coupling induces sfermion flavour mixing and 
a splitting of the third generation sfermion masses from the first two, 
due to RG effects from the Planck scale to the unification scale.
This applies also to the left-handed sleptons which have the mass
matrix in the flavour basis \cite{gabbi,barbi95};
\begin{equation} \label{slm}
 \tilde{m}^2_{ij}=m_0^2 \delta_{ij} - V^e_{3i}V^e_{3j}I_G
\end{equation}
where $m_0$ is the universal soft mass at the Planck scale and 
$I_G\lesssim m_0^2$ arises due to the large top Yukawa coupling.
%from the RG running to the unification scale.  
Here $V^e$ is related to the CKM matrix by
 $V^e_{31,32} \approx y V_{31,32}$ and 
 $V^e_{33} \approx V_{33}$ with $y \approx 0.7$.  Then, again in
the limit $m_0 \gg m_{\tilde{W}^-}$ we have 
\begin{eqnarray} 
&& \nonumber
 I^w_{12} \approx {\alpha_2\over 8\pi} y^2 \lambda^5 {I_G \over m_0^2},\quad
 I^w_{13} \approx {\alpha_2\over 8\pi} y \lambda^3 {I_G\over m_0^2} ,\\
&& 
 I^w_{23} \approx {\alpha_2\over 8\pi} y \lambda^2{I_G\over m_0^2} ,\quad
 I^w_{33}-I^w_{11} \approx {\alpha_2\over 8\pi} {I_G\over m_0^2}   \,, 
\end{eqnarray}
where $\lambda\approx 0.22$ is the Cabbibo angle. Note that even in such
a scenario the flavour diagonal elements 
$I^w_{33}-I^w_{11}$ and $\epsilon_\tau$ (for
$\tan\beta \gtrsim 10$)  give dominant contributions.
However, the departures from universality induced by the running from
the Planck scale to the unification scale may still play some role.
From the previous discussions, we can read the following effects on
each mass pattern set by the seesaw mechanism at $M$:
(i) With $m_1=m_2$, the fixed point relation $U_{31,32}=0$  follows from
the dominance of $\epsilon_\tau$ or 
$I^w_{33}-I^w_{11}\sim 10^{-3}$.
As $\Delta m^2 \sim m_1^2 \epsilon_\tau$ or $m_1^2(I^w_{33}-I^w_{11})$,
we cannot have the MSW solutions for reasonable values of $\tan\beta$.
On the other hand, the VO solution can be easily obtained.
With $m_1=-m_2$, it is allowed to have $U_{31}^2=U_{32}^2$ 
for which the contributions from $I^w_{33}-I^w_{11}$ and 
$\epsilon_\tau$ vanish [see Eq.~(\ref{diagbound})]. 
Then the VO solution can be obtained as
$\Delta m^2 \sim m_1^2 I^w_{23}$  and $I^w_{23}\sim 10^{-4}$ 
for $m_1^2\sim 10^{-6} eV^2$.
(ii) Contrary to case (i), we have $m_1^2 \sim 10^{-3} eV^2$ 
and thus only the MSW solutions can be realized.  Note also that
the relation $U_{31}^2=U_{32}^2$ with $m_1=-m_2$ yields an undesirable
result $\Delta m^2 \sim m_1^2 I^w_{23} \sim 10^{-7} eV^2$.
(iii) The dominance by $I^w_{33}-I^w_{11}$ or $\epsilon_\tau$ leads to
the situation, $\Delta m^2 \sim \Delta M^2$, with $|m_1|=|m_2|=m_3$.  
Thus it is not possible to generate two distinct mass-squared differences.
The only way to realize the cosmologically interesting possibility
$m_a \sim 1 eV$ is to have $\Delta M^2$ set by an initial texture with 
$m_1=-m_2 \neq m_3$ and $s_2=0$, $\sin^22\theta_3=1$ 
(implied by Eq.~(\ref{Mee})).  
Then, we get $U_{31}^2=U_{32}^2$ and $\Delta m^2 \sim m_1^2 I^w_{23}$ 
which is consistent with the LAMSW solution.  For this we need 
$\tan\beta \lesssim 15$ to suppress the $\epsilon_\tau$ contribution to 
$\Delta M^2$.  Note that the RG running of slepton masses from the 
the unification scale to the Majorana scale arising from 
Dirac neutrino Yukawa couplings \cite{borz} gives
only small corrections to the included effect in Eq.~(\ref{slm}).

\smallskip

In conclusion, we have discussed the effects of the flavour violating
slepton exchange on the neutrino masses and mixing patterns.
The requirement that this new contribution is consistent with
phenomenologically acceptable mass matrices without necessity of large
cancellation among different contributions gives for some mass patterns
interesting upper bounds on flavour violation in the slepton sector.
On the other hand, we find that for low and intermediate $\tan\beta$
values the slepton exchange contribution can be more important
than the usual tau Yukawa coupling effect included in RG evolution from 
the Majorana scale $M$ to the weak scale and can be
responsible for the observed neutrino masses and mixing.
For degenerate or partially degenerate mass textures at the Majorana scale,
the dominance of slepton exchange contribution gives new relations
among mixing angles, different from the RG fixed point solutions 
\cite{chan}. 
%In a certain theoretical framework like SO(10) unification,
%our considerations lead to more predictive results.
Finally we have discussed supersymmetric SO(10) unification as
an example of predictive frameworks for the slepton
flavour structure.

%\end{multicols}

%\newpage
\vspace{.3cm}

\begin{center}
FIG.~1. A neutrino self-energy diagram at one-loop involving
 left-handed slepton mixing.
\begin{picture}(500,100)(0,0)
\Vertex(150,10){2} 
\Vertex(250,10){2} 
\Line(197,56)(203,64)
\Line(197,64)(203,56)
\ArrowLine(100,10)(150,10)
\Text(125,20)[]{$\nu_i$}
\ArrowLine(150,10)(250,10)
\Text(200,0)[]{$\tilde{W}^-,\tilde{Z}$}
\ArrowLine(250,10)(300,10)
\Text(275,20)[]{$\nu_j$}
\DashArrowArcn(200,10)(50,180,90){3}
\Text(160,60)[]{$\tilde{l}_i$}
\DashArrowArcn(200,10)(50,90,0){4}
\Text(240,60)[]{$\tilde{l}_j$}
\end{picture}
\end{center}

\vspace{.3cm}

\begin{center}
%\begin{table}
%\caption{}
TABLE I. Mixing angle relations obtained for  dominant $I^w_{33}-
I^w_{11}$ or $I^w_{ij}$ ($i\neq j$) and tree-level mass degeneracies.

\begin{tabular}{c|ccc} 
 & $I_{12}=0\;(m_1=m_2)$ & $I_{13}=0\;(m_1=m_3)$ & $I_{23}=0\;(m_2=m_3)$  \\ 
             \hline
$I^w_{33}-I^w_{11}$ & &
   $\sin^22\theta_3={\sin^22\theta_1 s_2^2/(s_1^2+c_1^2s_2^2)^2}$ &
                \\ 
%$I^w_{12}$
%           & $s_2^2=\cot^2\theta_1\cot^22\theta_3$
%           & ${s_2^2\over (1-2s_2^2)^2}= \tan^2\theta_1\cot^2\theta_3$
%           & ${s_2^2\over (1-2s_2^2)^2}= \tan^2\theta_1\tan^2\theta_3$  \\
$I^w_{13}$
           & $s_2^2=\tan^2\theta_1\cot^22\theta_3$
           & ${s_2^2\over (1-2s_2^2)^2}= \cot^2\theta_1\cot^2\theta_3$
           & ${s_2^2\over (1-2s_2^2)^2}= \cot^2\theta_1\tan^2\theta_3$ \\
$I^w_{23}$
       & ${4s_2^2\over (1+s_2^2)^2}= \tan^22\theta_1\tan^22\theta_3$
           & $s_2^2=\cot^22\theta_1\tan^2\theta_3$
           & $s_2^2=\cot^22\theta_1\cot^2\theta_3$   \\ 
\end{tabular}
%\end{table}
\end{center}

\end{document}